\RequirePackage{ifpdf}
\ifpdf 
\documentclass[pdftex]{sigma}
\else
\documentclass{sigma}
\fi

\begin{document}

\allowdisplaybreaks

\renewcommand{\PaperNumber}{111}

\FirstPageHeading

\ShortArticleName{Second-Order Conformally Equivariant Quantization in Dimension $1|2$}

\ArticleName{Second-Order Conformally Equivariant Quantization\\ in Dimension $\boldsymbol{1|2}$}

\Author{Najla MELLOULI}

\AuthorNameForHeading{N. Mellouli}

\Address{Institut Camille Jordan, UMR 5208 du CNRS,
Universit\'e Claude Bernard Lyon 1,\\
43 boulevard du 11 novembre 1918,
69622 Villeurbanne cedex,
France}
\Email{\href{mailto:mellouli@math.univ-lyon1.fr}{mellouli@math.univ-lyon1.fr}}

\ArticleDates{Received September 22, 2009, in f\/inal form December 13, 2009; Published online December 28, 2009}

\Abstract{This paper is the next step of an ambitious program to develop
conformally equivariant quantization on supermanifolds. This problem
was considered so far in (super)dimensions 1 and  $1|1$. We will
show that the case of several odd variables is much more dif\/f\/icult.
We consider the supercircle $S^{1|2}$ equipped with the standard
contact structure. The conformal Lie superalgebra ${\mathcal{K}}(2)$
of contact vector f\/ields on $S^{1|2}$ contains the Lie superalgebra
$\mathrm{osp}(2|2)$. We study the spaces of linear dif\/ferential
operators on the spaces of weighted densities as modules over
$\mathrm{osp}(2|2)$. We prove that, in the non-resonant case, the
spaces of second order dif\/ferential operators are isomorphic to the
corresponding spaces of symbols as $\mathrm{osp}(2|2)$-modules. We
also prove that the conformal equivariant quantization map is unique
and calculate its explicit formula.}


\Classification{17B10; 17B68; 53D55}

\section{Introduction and the main results}

The concept of equivariant quantization f\/irst appeared in \cite{LO} and \cite{DLO}. The general idea is to identify, in a canonical way, the space of
linear dif\/ferential operators on a manifold acting on weighted densities
with the corresponding space of symbols. Such an identif\/ication is called a~quantization (or symbol) map. It turns out that for an arbitrary
projectively/conformally f\/lat manifold, there exists a unique quantization
map commuting with the action of the group of projective/conformal
transformations.

Equivariant quantization on supermanifolds was initiated by~\cite{CMZ} and
further investigated in~\cite{GMO}. In these works, the authors considered
supermanifolds of dimension $1|1$. This is in part due to the fact that, in
the super cases considered, one has to take into account a non-integrable
distribution, namely the contact structure, see~\cite{Lei,GLS1}, dubbed
``SUSY'' structure in \cite{CMZ}.

In this paper, we consider the space of linear dif\/ferential operators on the
supercircle~$S^{1|2}$ acting from the space of $\lambda$-densities to the
space of $\mu$-densities, where $\lambda$ and~$\mu$ are (real or complex)
numbers. This space of operators is naturally a module over the Lie
superalgebra of contact vector f\/ields (see Section \ref{XXX} below) also
known as the stringy superalgebra $\mathcal{K}(2)$, see \cite{GLS1}. We
denote these modules by $\mathcal{D}_{\lambda ,\mu}\left( S^{1|2}\right)$.

Our main result concerns the spaces containing second-order dif\/ferential
operators, $\mathcal{D}_{\lambda \mu }^{\frac{3}{2}}\left( S^{1|2}\right) $
and $\mathcal{D}_{\lambda \mu }^{2}\left( S^{1|2}\right) $. The space $%
\mathcal{D}_{\lambda \mu }^{\frac{3}{2}}\left( S^{1|2}\right) $ is contained
in $\mathcal{D}_{\lambda \mu }^{2}\left( S^{1|2}\right) $, so we will be
interested to the space $\mathcal{D}_{\lambda \mu }^{2}\left( S^{1|2}\right)
$. This space is not isomorphic to the corresponding space of symbols, as a~$\mathcal{K}(2)$-module. The obstructions to the existence of such an
isomorphism are given by (the inf\/initesimal version of) the Schwarzian
derivative, see \cite{MD} and references therein. We thus restrict the
module structure on $\mathcal{D}_{\lambda \mu }^{2}\left( S^{1|2}\right) $
to the orthosymplectic Lie superalgebra $\mathrm{osp}\left( 2|2\right) $
naturally embedded to $\mathcal{K}(2)$.

The main result of this paper is as follows.

\begin{theorem}
\label{mainthm} $(i)$ The space $\mathcal{D}_{\lambda \mu }^{2}\left(
S^{1|2}\right) $ and the corresponding space of symbols are isomorphic as $\mathrm{osp}\left( 2|2\right) $-modules, provided $\mu -\lambda \neq 0,\frac{1}{2},1,\frac{3}{2},2$.

$(ii)$ The above isomorphism is unique.
\end{theorem}

The particular values of $\lambda$ and $\mu$ such that $%
\mu-\lambda\in\{0, \frac{1}{2}, 1,\frac{3}{2},2\}$ are called \textit{%
resonant}. We do not study here the corresponding ``resonant modules'' of
dif\/ferential operators. Note that these modules are of particular interest
and deserve further study.

We think that a similar result holds for the space of dif\/ferential
operators of arbitrary order, but such a result is out of reach so
far. We would like to mention however, that most of the known
interesting examples of dif\/ferential operators in geometry and
mathematical physics are of order 2. This allows one to expect
concrete applications of the above theorem.

\section[Geometry of the supercircle $S^{1|2}$]{Geometry of the supercircle $\boldsymbol{S^{1|2}}$}

The supercircle $S^{1|2}$ is a supermanifold of dimension $1|2$ which
generalizes the circle $S^{1}$. To f\/ix the notation, let us give the basic
def\/initions of geometrical objects on $S^{1|2}$, see~\cite{Lei,GLS1} for
more details.

We def\/ine the supercircle $S^{1|2}$ by describing its graded commutative
algebra of (complex valued) functions that we note by $C^{\infty }\left(
S^{1|2}\right) $, consisting of the elements
\begin{gather*}
f\left( x,\xi _{1},\xi _{2}\right) =  f_{0}(x)+\xi _{1} f_{1}(x)+\xi
_{2} f_{2}(x)+\xi _{1}\xi _{2} f_{12}(x),  
\end{gather*}
where $x$ is the Fourier image of the angle parameter on $S^{1}$ and $\xi
_{1}$, $\xi _{2}$ are odd Grassmann coordinates, i.e., $\xi _{i}^{2}=0$, $\xi
_{1}\xi _{2}=-\xi _{2}\xi _{1}$ and where $f_{0},f_{12},f_{1},f_{2}\in
{}C^{\infty }(S)$ are functions with complex values. We def\/ine the parity
function~$p$ by setting $p\left( x\right) =0$ and $p\left( \xi _{i}\right)
=1 $.

\subsection{Vector f\/ields and dif\/ferential forms}

Any \textit{vector field on} $S^{1|2}$ is a derivation of the algebra $%
C^{\infty }\left( S^{1|2}\right)$, it can be expressed as
\begin{gather*}
X=f\partial _{x}+g_{1}\partial _{\xi _{1}} +g_{2}\partial _{\xi _{2}}
\end{gather*}
with $f,g_i\in C^{\infty }\left( S^{1|2}\right)$, $\partial _{x}=\frac{
\partial }{\partial x}$ and $\partial _{\xi _{i}}=\frac{\partial }{\partial
\xi _{i}}$, for $i=1,2.$ The space of vector f\/ields on $S^{1|2}$ is a Lie
superalgebra which we note by $\mathrm{Vect} \left( S^{1|2}\right)$.

Any \textit{differential form} is a skew-symmetric multi-linear map (over $%
C^{\infty }\left( S^{1|2}\right)$) from $\mathrm{Vect} \left(S^{1|2}\right)$ to $C^{\infty }\left( S^{1|2}\right)$. To f\/ix the notation,
we set $\left\langle \partial _{u_{i}},du_{j}\right\rangle =\delta_{ij}$,
for $u=\left( x,\xi _{1},\xi_{2}\right)$. The space of dif\/ferential forms $%
\Omega ^{1}\left( S^{1|2}\right)$ is a right $C^{\infty}\left(
S^{1|2}\right)$-module and a left $\mathrm{Vect} \left( S^{1|2}\right)$%
-module, the action being given by the Lie derivative, i.e., $\left\langle
X,L_{Y}\alpha \right\rangle :=\left\langle \left[ X,Y\right] ,\alpha
\right\rangle $ for any $Y,X\in\mathrm{Vect} \left( S^{1|2}\right)$,
$\alpha \in \Omega ^{1}\left( S^{1|2}\right)$.

\subsection{The Lie superalgebra of contact vector f\/ields}

The \textit{standard contact} structure on $S^{1|2}$ is def\/ined by the data
of a linear distribution $\left\langle \overline{D}_{1},\overline{D}
_{2}\right\rangle $ on $S^{1|2}$ generated by the odd vector f\/ields
\begin{gather*}  
\overline{D}_{1}=\partial _{\xi _{1}}-\xi _{1}\partial _{x}, \qquad\overline{
D}_{2}=\partial _{\xi _{2}}-\xi _{2}\partial _{x}.
\end{gather*}
This contact structure can also be def\/ined as the kernel of the dif\/ferential
$1$-form:
\begin{gather*}
\alpha =dx+\xi _{1}d\xi _{1}+\xi _{2}d\xi _{2}.
\end{gather*}
We refer to \cite{Shch} for more details.

A vector f\/ield $X$ on $S^{1|2}$ is called a \textit{contact vector field} if
it preserves the contact distribution, that is, satisf\/ies the condition:
\begin{gather*}
\left[ X,\overline{D}_{1}\right] =\psi _{1_{X}} \overline{D}_{1}+\psi
_{2_{X}}  \overline{D}_{2}, \qquad \left[ X,\overline{D}_{2}\right] = \phi
_{1_{X}} \overline{D}_{1}+ \phi _{2_{X}} \overline{D}_{2},
\end{gather*}
where $\psi _{1_{X}},\psi _{2_{X}},\phi _{1_{X}},\phi _{2_{X}}\in
C^{\infty }\left( S^{1|2}\right)$ are functions depending on $X$. The
space of the contact vector f\/ields is a Lie superalgebra which we note by $\mathcal{K}\left(2\right)$.

The following fact is well-known.

\begin{lemma}
Every contact vector field $($see~{\rm \cite{Lei}}$)$ can be expressed, for some
function $f\in C^{\infty }\!\left( S^{1|2}\right)$, by
\begin{gather}
X_{f}=f\partial _{x}-\left( -1\right) ^{p\left( f\right) }\tfrac{1}{2}\left(
\overline{D}_{1}\left( f\right) \overline{D}_{1}  +\overline{D}_{2}\left(
f\right) \overline{D}_{2}\right) . \label{ContVF}
\end{gather}
\end{lemma}

The function $f$ is said to be a \textit{contact Hamiltonian} of the f\/ield $%
X_{f}$. The space $C^{\infty }\left( S^{1|2}\right) $ is therefore
identif\/ied with the Lie superalgebra $\mathcal{\ K}\left( 2\right) $ and
equipped with the structure of Lie superalgebra with respect to the contact
bracket:
\begin{gather}
\left\{ f,g\right\} = fg^{\prime }-f^{\prime }g-\left( -1\right) ^{p\left(
f\right) }\tfrac{1}{2} \left( \overline{D}_{1}\left( f\right) \overline{D}
_{1}\left( g\right)   +\overline{D} _{2}\left( f\right) \overline{D}
_{2}\left( g\right) \right),  \label{CoP}
\end{gather}
where $f^{\prime }=\partial _{x}(f)$.

\subsection{Conformal symmetry: the orthosymplectic superalgebra}

The \textit{conformal} (or projective) structure on the supercircle $S^{1|2}$
(see \cite{OOC}) is def\/ined by the action of the $4|4$-dimensional Lie
superalgebra $\mathrm{osp}\left(2|2\right)$. This action is spanned by the
contact vector f\/ields with the contact Hamiltonians:
\[
\left\{ 1,x,x^2,\xi_1\xi_2;\,\xi_1,\xi_2,x\xi_1, x\xi_2 \right\}.
\]
The embedding of $\mathrm{osp}\left(2|2\right)$ into $\mathcal{K}
\left(2\right)$ is then given by~(\ref{ContVF}).

The subalgebra $\mathrm{Af\/f}\left(2|2\right)$ of $\mathrm{osp}
\left(2|2\right)$ spanned by the contact vector f\/ields with the contact
Hamiltonians $\left\{ 1,x,\xi_1\xi_2;\, \xi_1,\xi_2 \right\} $ will be called
the \textit{affine} Lie superalgebra.

\subsection{Modules of weighted densities}

We introduce a family of $\mathcal{K}$ $\left( 2\right) $-modules with a
parameter. For any contact vector f\/ield, we def\/ine a family of
dif\/ferential operators of order one on $C^{\infty }\left(
S^{1|2}\right) $
\begin{gather}
L_{X_{f}}^{\lambda }:=X_{f}+\lambda f^{\prime },  \label{LieDer}
\end{gather}
where the parameter $\lambda $ is an arbitrary (complex) number and the
function is considered as a~0-order dif\/ferential operator of left
multiplication by this function. The map $X_{f}\mapsto L_{X_{f}}^{\lambda }$
is a~homomorphism of Lie superalgebras. We thus obtain a family of $\mathcal{K}\left( 2\right) $-modules on $C^{\infty }\left( S^{1|2}\right) $ that
we note by $\mathcal{F}_{\lambda }\left( S^{1|2}\right) $ and that we call
spaces of \textit{weighted densities} of weight $\lambda $.

Viewed as vector spaces, but not as $\mathcal{K}\left( 2\right)$-modules,
the spaces $\mathcal{F}_{\lambda }\left( S^{1|2}\right)$ are isomorphic to $C^{\infty }\left( S^{1|2}\right)$.

The space of weighted densities possesses a Poisson superalgebra structure
with respect to the contact bracket $\{\cdot,\cdot\}:\mathcal{F}_{\lambda }\left(
S^{1|2}\right) \otimes \mathcal{F}_{\mu }\left( S^{1|2}\right) \rightarrow
\mathcal{F}_{\lambda +\mu +1}\left( S^{1|2}\right) $ given explicitly by
\begin{gather*}
\left\{ f,g\right\} =\mu  f^{\prime }g-\lambda  fg^{\prime }-\left(
-1\right) ^{p\left( f\right) }\tfrac{1}{2}\left( \overline{D}_{1}\left(
f\right) \overline{D}_{1}\left( g\right) +\overline{D}_{2}\left( f\right)
\overline{D}_{2}\left( g\right) \right) 
\end{gather*}
for all $f\in \mathcal{F}_{\lambda }\left( S^{1|2}\right)$, $g\in \mathcal{F}_{\mu }\left( S^{1|2}\right) .$

Note that if $f\in\mathcal{F}_{-1}\left( S^{1|2}\right)$, then $\left\{
f,g\right\} =L_{X_{f}}^{\mu }(g)$. The space $\mathcal{F}_{-1}\left(
S^{1|2}\right)$ is a subalgebra isomorphic to $\mathcal{K}\left(2\right)$,
see formula~(\ref{CoP}).

\section{Dif\/ferential operators on the spaces of weighted densities}
\label{XXX}

In this section we introduce the space of dif\/ferential operators acting on
the spaces of weighted densities and the corresponding space of symbols on $%
S^{1|2}$. We refer to \cite{LKV,GLS,GMO,Con} for further details. This space
is naturally a module over the Lie superalgebra $\mathcal{K}(2)$.

We also def\/ine a $\mathcal{K}(2)$-invariant ``f\/iner f\/iltration'' on the
modules of dif\/ferential operators that plays the key role in this paper. The
graded $\mathcal{K}(2)$-module associated to the f\/iner f\/iltration is called
the \textit{module of symbols}.

\subsection[Definition of the modules $\mathcal{D}^{(k)}_{\lambda\mu}\left( S^{1|2}\right)$]{Def\/inition of the modules $\boldsymbol{\mathcal{D}^{(k)}_{\lambda\mu}\left( S^{1|2}\right)}$}

Let $\mathcal{D}_{\lambda\mu }\left( S^{1|2}\right)$ be the space of linear
dif\/ferential operators $A:\mathcal{F}_{\lambda }\left( S^{1|2}\right)
\rightarrow \mathcal{F}_{\mu }\left( S^{1|2}\right)$.

This space is naturally f\/iltered:
\[
\mathcal{D}_{\lambda \mu }^{(0)}\big( S^{1|2}\big) \subset \mathcal{D}
_{\lambda \mu }^{(1)}\big( S^{1|2}\big) \subset\cdots\subset \mathcal{D}
_{\lambda \mu }^{(k-1)}\big( S^{1|2}\big) \subset \mathcal{D}_{\lambda
\mu }^{(k)}\big( S^{1|2}\big)\subset\cdots,
\]
where $\mathcal{D}_{\lambda \mu }^{(k)}\left( S^{1|2}\right)$ is the space
of linear dif\/ferential operators of order $k$.

The space $\mathcal{D}_{\lambda\mu }\left( S^{1|2}\right)$ and every
subspace $\mathcal{D}_{\lambda \mu }^{(k)}\left( S^{1|2}\right)$ is
naturally a module over the Lie superalgebra of contact vector f\/ields $%
\mathcal{K}\left(2\right)$. The above f\/iltration is of course $\mathcal{K}
\left(2\right)$-invariant. Note that, in the case $\lambda=\mu=0$, the space
of dif\/ferential operators is a module over the full Lie superalgebra $%
\mathrm{Vect}\left( S^{1|2}\right)$ and the above f\/iltration is $%
\mathrm{Vect}\left( S^{1|2}\right)$-invariant.

\subsection[The finer filtration: modules $\mathcal{D}^k_{\lambda\mu}\left( S^{1|2}\right)$]{The f\/iner f\/iltration: modules $\boldsymbol{\mathcal{D}^k_{\lambda\mu}\left( S^{1|2}\right)}$}

It turns out that there is another, f\/iner f\/iltration:
\begin{gather}
\mathcal{D}_{\lambda \mu }^{0}\big( S^{1|2}\big) \subset \mathcal{D}%
_{\lambda \mu }^{\frac{1}{2}}\big( S^{1|2}\big)
\subset \mathcal{D}%
_{\lambda \mu }^{1}\big( S^{1|2}\big) \subset \mathcal{D}_{\lambda \mu }^{%
\frac{3}{2}}\big( S^{1|2}\big)
 \subset \mathcal{D}_{\lambda \mu }^{2}\big(
S^{1|2}\big) \subset \cdots  \label{finfil}
\end{gather}
on the space of dif\/ferential operators on $S^{1|2}$. This f\/iner f\/iltration
is invariant with respect to the action of $\mathcal{K}\left( 2\right) $
(but it cannot be invariant with respect to the action of the full algebra
of vector f\/ields).

\begin{proposition}
Every differential operator can be expressed in the form
\begin{gather}
A=\sum_{\ell ,m,n}a_{\ell ,m,n}\left( \partial _{x}\right) ^{\ell } %
\overline{D}_{1}^{m} \overline{D}_{2}^{n},  \label{General}
\end{gather}
where $a_{\ell ,m,n}\in C^{\infty }\left( S^{1|2}\right) $, the index $%
\ell $ is arbitrary while $m,n\leq 1$, and where only finitely many terms
are non-zero.
\end{proposition}

\begin{proof}
If $A\in\mathcal{D}_{\lambda \mu }\left( S^{1|2}\right)$, then
$
A=\sum
a_{\ell,m,n} \left(\partial _{x}\right)^\ell
\partial _{\xi _{1}}^m \partial _{\xi _{2}}^n
$
and since one has:
\[
\partial _{\xi _{1}}=\overline{D}_{1}+\xi _{1}\partial _{x},
\qquad
\partial _{\xi _{2}}=\overline{D}_{2}+\xi _{2}\partial _{x},
\qquad
\overline{D}_i^2=-\partial _{x},
\]
we have the form desired.
\end{proof}

For every (half)-integer $k$, we denote by $\mathcal{D}_{\lambda \mu
}^{k}\left( S^{1|2}\right) $ the space of dif\/ferential operators of the form
\begin{gather}
A=\sum_{\ell +\frac{m}{2}+\frac{n}{2}\leq  k}a_{\ell ,m,n}\left( \partial
_{x}\right) ^{\ell } \overline{D}_{1}^{m} \overline{D}_{2}^{n},
\label{GenForm}
\end{gather}
where $a_{\ell ,m,n}\in C^{\infty }\left( S^{1|2}\right) .$
Furthermore, since $\partial _{x}=-\overline{D}_{1}^{2}=-\overline{D}%
_{2}^{2} $, we can assume $m,n\leq 1$.

\begin{proposition}
The form \eqref{GenForm} is stable with respect to the action of $\mathcal{K}
\left(2\right)$.
\end{proposition}

\begin{proof}
Let $X_f$ be a contact vector f\/ield, see formula (\ref{ContVF}).
The action of $X_f$ on the space
$\mathcal{D}_{\lambda \mu }\left( S^{1|2}\right)$ is given by
\begin{gather}
\label{GenLD}
{\cal L}_{X_f}(A)=
L^\mu_{X_f}\circ{}A-(-1)^{p(f)p(A)} A\circ{}L^\lambda_{X_f},
\end{gather}
where $L^\lambda_{X_f}$ is the Lie derivative (\ref{LieDer}). The
invariance of the form (\ref{GenForm}) is subject to a
straightforward calculation.
\end{proof}

\begin{remark} 1) It is worth noticing that for $k$ integer, one has
\[
\mathcal{D}_{\lambda \mu }^{(k)}\big( S^{1|2}\big) \subset
 \mathcal{D}_{\lambda \mu
}^{k}\big( S^{1|2}\big)
\]
but these modules do not coincide. Indeed, the
module $\mathcal{D}_{\lambda \mu }^{k}\left( S^{1|2}\right) $ contains
operators proportional to $\partial _{x}^{k-1} \overline{D}_{1} \overline{D%
}_{2}$ which are, of course, of order $k+1$. An element of $\mathcal{D}_{\lambda \mu }^{k}\left( S^{1|2}\right) $ will be called $k$-dif\/ferential
operator, it does not have to be of order~$k$, it can be of order $k+1$.

2) A similar f\/iner f\/iltration exists for an arbitrary contact
manifold, cf.~\cite{Ovs,FMP}. In the $1|1$-dimensional case this f\/iner
f\/iltration was used in~\cite{GMO}.
\end{remark}

\begin{example}
The module $\mathcal{D}_{\lambda \mu }^{2}\left( S^{1|2}\right) $ will
be particularly interesting for us. Every dif\/ferential operator $A\in
\mathcal{D}_{\lambda \mu }^{2}\left( S^{1|2}\right) $ can be expressed in
the form
\begin{gather*}
A   =   a_{0,0,0}   +  a_{0,1,0}\overline{D}_{1}  +  a_{0,0,1}\overline{D}_{2}
 +  a_{1,0,0}\partial _{x}  +  a_{0,1,1}\overline{D}_{1}\overline{D}
_{2}  +  a_{1,1,0}\partial _{x}\overline{D}_{1}\\
\phantom{A=}{}  +  a_{1,0,1}\partial _{x}
\overline{D}_{2}  +  a_{2,0,0}\partial _{x}^{2}  +  a_{1,1,1}\partial _{x}\overline{D}
_{1}\overline{D}_{2}.
\end{gather*}
\end{example}

\subsection{Space of symbols of dif\/ferential operators}

We consider the graded $\mathcal{K}\left( 2\right)$-module associated to the
f\/ine f\/iltration (\ref{finfil}):
\begin{gather*}  
\mathrm{gr}\,\mathcal{D}_{\lambda \mu}\big( S^{1|2}\big) =
\bigoplus_{i=0}^\infty \mathrm{gr}^{\frac{i}{2}}\mathcal{D}_{\lambda
\mu}\big( S^{1|2}\big),
\end{gather*}
where $\mathrm{gr}^k \mathcal{D}_{\lambda \mu}\left( S^{1|2}\right)=
\mathcal{D}_{\lambda \mu }^{k}\left( S^{1|2}\right)/ \mathcal{D}_{\lambda
\mu }^{k-\frac{1}{2}}\left( S^{1|2}\right)$ for every (half)integer $k$.
This module is called the \textit{space of symbols} of dif\/ferential
operators.

The image of a dif\/ferential operator $A$ under the natural projection
\[
\sigma _{\rm pr}: \ \mathcal{D}_{\lambda \mu }^{k}\big( S^{1|2}\big) \rightarrow
\mathrm{gr}^{k} \mathcal{D}_{\lambda \mu }\big( S^{1|2}\big)
\]
def\/ined by the f\/iltration (\ref{finfil}) is called the \textit{principal
symbol}.

We need to know the action of the Lie superalgebra $\mathcal{K}
\left(2\right)$ on the space of symbols.

\begin{proposition}
\label{SymP} If $k$ is an integer, then
\begin{gather*}
\mathrm{gr}^k\mathcal{D}_{\lambda \mu}\big( S^{1|2}\big) =\mathcal{F}
_{\mu -\lambda -k} \oplus \mathcal{F}_{\mu -\lambda -k}
\end{gather*}
\end{proposition}

\begin{proof}
By def\/inition (see formula (\ref{GenForm})), a given operator
$A\in\mathcal{D}^k_{\lambda \mu}\left( S^{1|2}\right)$
with integer $k$ is of the form
\[
A=F_1 \partial_x^k+F_2 \partial_x^{k-1}\overline{D}_1\overline{D}_2+\cdots,
\]
where $\cdots$ stand for lower order terms.
The principal symbol of $A$ is then encoded by the pair $(F_1,F_2)$.
From~(\ref{GenLD}), one can easily calculate the $\mathcal{K}(2)$-action on
the principal symbol:
\[
L_{X_{f}}\left(
F_{1},
F_{2}
\right) =\big(
L_{X_{f}}^{\mu -\lambda -k}\left( F_{1}\right),
L_{X_{f}}^{\mu -\lambda -k}\left( F_{2}\right)
\big).
\]
In other words, both $F_1$ and $F_2$ transform as
$(\mu -\lambda -k)$-densities.
\end{proof}

Surprisingly enough, the situation is more complicated in the case of
half-integer $k$.

\begin{proposition}
\label{SymPBis} If $k$ is a half-integer, then the $\mathcal{K}\left(
2\right) $-action is as follows:
\begin{gather}
L_{X_{f}}(F_{1},F_{2})=\left(L_{X_{f}}^{\mu -\lambda -k}\left( F_{1}\right) -%
\tfrac{1}{2} \overline{D}_{1} \overline{D}_{2}\left( f\right)
F_{2}, L_{X_{f}}^{\mu -\lambda -k}\left( F_{2}\right) +\tfrac{1}{2}\overline{%
D}_{1} \overline{D}_{2}\left( f\right) F_{1}\right).  \label{Sact}
\end{gather}
\end{proposition}

\begin{proof}
Directly from (\ref{GenLD}).
\end{proof}

This means that the space of symbols of half-integer contact order is not isomorphic
to the space of weighted densities. It would be nice to understand the
geometric nature of the action~(\ref{Sact}).

\begin{corollary}
The module $\mathrm{gr}\, \mathcal{D}_{\lambda \mu}\left(
S^{1|2}\right)$ depends only on $\mu -\lambda$.
\end{corollary}

\noindent Following \cite{LO,DLO,GMO} and to simplify the notation, we will
denote by $\mathcal{S}_{\mu -\lambda }\left( S^{1|2}\right) $ the full space
of symbols $\mathrm{gr}\mathcal{D}_{\lambda \mu }\left( S^{1|2}\right) $ and
$\mathcal{S}_{\mu -\lambda }^{k}\left( S^{1|2}\right) $ the space of symbols
of contact order $k$.

\begin{corollary}
The space $\mathcal{D}_{\lambda \mu }^{k}\left( S^{1|2}\right) $ is
isomorphic to $\mathcal{S}_{\mu -\lambda }^{k}\left( S^{1|2}\right)
$ as a module over the affine Lie superalgebra $\mathrm{Af\/f}\left(
2|2\right) $.
\end{corollary}

\begin{proof}
To def\/ine an $\mathrm{Af\/f}\left(2|2\right)$-equivariant quantization map,
it suf\/f\/ice to consider the inverse of the principal symbol:
$Q=\sigma_{\rm pr}^{-1}$.
\end{proof}

A linear map $Q:\mathcal{S}_{\mu -\lambda }\left( S^{1|2}\right) \rightarrow
\mathcal{D}_{\lambda \mu }\left( S^{1|2}\right)$ is called a \textit{quantization map} if it is bijective and preserves the principal symbol of
every dif\/ferential operator, i.e., $\sigma_{\rm pr}\circ{}Q=\mathrm{Id}$. The
inverse map $\sigma =Q^{-1}$ is called a \textit{symbol map}.

\section[Conformally equivariant quantization on $S^{1|2}$]{Conformally equivariant quantization on $\boldsymbol{S^{1|2}}$}

In this section we prove the main results of this paper~-- Theorem~\ref{mainthm}~-- on the existence and uniqueness of the conformally equivariant
quantization map on the space $\mathcal{D}_{\lambda \mu }^{2}\left(
S^{1|2}\right) $. We calculate this quantization map explicitly.

We already proved that the space $\mathcal{D}_{\lambda \mu }^2\left(
S^{1|2}\right)$ is isomorphic to the corresponding space of symbols as a
module over the af\/f\/ine Lie superalgebra $\mathrm{Af\/f}\left(2|2\right)$. We
will now show how to extend this isomorphism to that of the $\mathrm{osp}%
\left( 2|2\right) $-modules.

\subsection[Equivariant quantization map in the case of $1/2$-differential operators]{Equivariant quantization map in the case of $\boldsymbol{\frac{1}{2}}$-dif\/ferential operators}

Let us f\/irst consider the quantization of symbols of $\frac{1}{2}$%
-dif\/ferential operators. By linearity, we can assume that the symbols of
dif\/ferential operators are homogeneous (purely even or purely odd). Since
for any symbol $\left(F_{1},F_{2}\right) \in \mathcal{S}_{\mu -\lambda }^{\frac{1}{2}}\left( S^{1|2}\right)$, we have $p(F_{1})=p(F_{2})$, we can
def\/ine \textit{parity of the symbol} $(F_{1},F_{2})$ as $%
p(F):=p(F_{1})=p(F_{2})$.

\begin{proposition}
The unique $\mathrm{osp}\left( 2|2\right)$-equivariant quantization
map
associates the following $\frac{1}{2}$-differential operator to a symbol $%
\left( F_{1},F_{2}\right)\in \mathcal{S}^{\frac{1}{2}}_{\mu -\lambda }\left(
S^{1|2}\right)$ provided $\mu\neq\lambda$:
\begin{gather*}
Q\left( F_{1},F_{2}\right) = F_{1}\overline{D}_{1}+ F_{2}\overline{D}_{2}+
(-1)^{p(F)} \frac{\lambda }{\lambda-\mu} \left(\overline{D}_{1}\left(
F_{1}\right) +\overline{D}_{2}\left( F_{2}\right) \right).
\end{gather*}
\end{proposition}

\begin{proof}
First, one easily checks that this quantization map is, indeed, $\mathrm{osp}\left( 2|2\right)$-equivariant. Let us prove the uniqueness.

Consider f\/irst an arbitrary dif\/ferentiable linear map $Q:\mathcal{S}_{\mu
-\lambda }^{\frac{1}{2}}\left( S^{1|2}\right) \rightarrow \mathcal{D}%
_{\lambda \mu }^{\frac{1}{2}}\left( S^{1|2}\right) $ preserving the
principal symbol. Such a map is of the form:
\[
Q\left( F_{1},F_{2}\right) =F_{1}\overline{D}_{1}+F_{2}\overline{D}_{2}+%
\widetilde{Q}_{1}^{\left( 1\right) }\left( F_{1}\right) +\widetilde{Q}%
_{2}^{\left( 1\right) }\left( F_{2}\right),
\]
where $\widetilde{Q}_{1}^{\left( 1\right) }$ and $\widetilde{Q}_{2}^{\left(
1\right) }$ are dif\/ferential operators with coef\/f\/icients in $\mathcal{F}%
_{\mu -\lambda }$, cf.\ formula (\ref{General}).

One then easily checks the following:

$a)$ This map commutes with the action of the vector f\/ields $D_{1},D_{2}\in
\mathrm{osp}\left( 2|2\right) $, where $D_{i}=\partial _{\xi _{i}}+\xi
_{i}\partial _{x}$, if and only if the dif\/ferential operators $\widetilde{Q}%
_{1}^{\left( 1\right) }$ and $\widetilde{Q}_{2}^{\left( 1\right) }$ are with
constant coef\/f\/icients.

$b)$ This map commutes with the linear vector f\/ields $X_{x}$, $X_{\xi
_{1}}$, $X_{\xi _{2}}$ if and only if the dif\/ferential operators $\widetilde{Q}%
_{1}^{\left( 1\right) }$ and $\widetilde{Q}_{2}^{\left( 1\right) }$ are of
order 1 and moreover have the form:
\[
\widetilde{Q}_{1}^{\left( 1\right) }\left( F_{1}\right) =C_{11}
\overline{D}_{1}\left( F_{1}\right), \qquad \widetilde{Q}_{2}^{\left(
1\right) }\left( F_{2}\right) =C_{12} \overline{D}_{2}\left(
F_{2}\right),
\]
where $C_{11}$, $C_{12}$ are arbitrary constants.

We thus determined the general form of a quantization map commuting with the
action of the af\/f\/ine subalgebra $\mathrm{Af\/f}\left( 2|2\right) $.

$c)$ This map commutes with $X_{\xi _{1}\xi _{2}}$ if and only if $%
C_{11}=C_{12}.$

In order to satisfy the full condition of $\mathrm{osp}\left( 2|2\right)$-equivariance, it remains to impose the equiva\-riance with respect to the
vector f\/ield $X_{x^{2}}$.

$d)$ The above quantization map commutes with the action of $X_{x^{2}}$ if and
only if $C_{11}$, $C_{12}$ satisfy the following condition:
\begin{gather*}
2\left( \mu -\lambda -\tfrac{1}{2}\right) C_{11}+C_{12}   =
(-1)^{p(F)+1} 2\lambda , \\
C_{11}+2\left( \mu -\lambda -\tfrac{1}{2}\right) C_{12}   =
(-1)^{p(F)+1} 2\lambda
\end{gather*}
If $\mu -\lambda \neq 0$, this system can be easily solved and the solution
is $C_{11}=C_{12}=(-1)^{p(F)}\frac{\lambda }{\lambda -\mu }$.
\end{proof}

\subsection{Equivariant quantization map in the case of 1-dif\/ferential
operators}

Let us consider the next case. All the calculations are similar (yet more
involved) to the above calculations.

\begin{proposition}
\label{FoDPr} The unique $\mathrm{osp}\left( 2|2\right)
$-equivariant
quantization map associates the following differential operator to a symbol $%
\left( F_{1},F_{2}\right) \in \mathcal{S}_{\mu -\lambda }^{1}\left(
S^{1|2}\right) $:
\begin{gather}
Q\left( F_{1},F_{2}\right)   =   F_{1}\partial _{x}+F_{2}\overline{D}_{1}%
\overline{D}_{2}
  +(-1)^{p(F)}\frac{1}{2\left( \mu -\lambda -1\right) }\left(  \overline{%
D}_{1}\left( F_{1}\right) \overline{D}_{1}+ \overline{D}_{2}\left(
F_{1}\right) \overline{D}_{2}\right) \nonumber\\
\phantom{Q\left( F_{1},F_{2}\right)   =}{}
 +(-1)^{p(F)}\frac{\lambda +\frac{1}{2}}{\mu -\lambda -1}\left(
\overline{D}_{2}\left( F_{2}\right) \overline{D}_{1}- \overline{D}%
_{1}\left( F_{2}\right) \overline{D}_{2}\right) \nonumber\\
\phantom{Q\left( F_{1},F_{2}\right)   =}{}
-\frac{\lambda }{\mu -\lambda -1}\left(  \partial _{x}\left(
F_{1}\right) -\frac{1+2\lambda }{2\left( \mu -\lambda \right) -1} \overline{%
D}_{1}\overline{D}_{2}\left( F_{2}\right) \right)
\label{FoQ1}
\end{gather}
provided $\mu -\lambda \neq \frac{1}{2},1$.
\end{proposition}

\begin{proof}
First, we check by a straightforward calculation that an arbitrary $\mathrm{%
Af\/f}\left( 2|2\right) $-equivariant quantization map is given by
\begin{gather*}
Q\left( F_{1},F_{2}\right)    =   F_{1}\partial _{x}+F_{2}\overline{D}_{1}%
\overline{D}_{2}
  +\widetilde{Q}_{1}^{\left( 1\right) }\left( F_{1}\right) +\widetilde{Q}%
_{2}^{\left( 1\right) }\left( F_{2}\right)
  +\widetilde{Q}_{1}^{\left( 2\right) }\left( F_{1}\right) +\widetilde{Q}%
_{2}^{\left( 2\right) }\left( F_{2}\right),
\end{gather*}
where the dif\/ferential operators $\widetilde{Q}_{1}^{\left( n\right) }$ and $\widetilde{Q}_{2}^{\left( n\right) }$
are of order $n$ and have the form:
\begin{gather*}
\widetilde{Q}_{1}^{\left( 1\right) }\left( F_{1}\right)    =   C_{11}%
\overline{D}_{1}\left( F_{1}\right) \overline{D}_{1}+C_{13}\overline{D}%
_{2}\left( F_{1}\right) \overline{D}_{2}, \\
\widetilde{Q}_{1}^{\left( 2\right) }\left( F_{1}\right)    =
C_{21}\partial _{x}\left( F_{1}\right),  \\
\widetilde{Q}_{2}^{\left( 1\right) }\left( F_{2}\right)    =   C_{12}%
\overline{D}_{2}\left( F_{2}\right) \overline{D}_{1}+C_{14}\overline{D}%
_{1}\left( F_{2}\right) \overline{D}_{2}, \\
\widetilde{Q}_{2}^{\left( 2\right) }\left( F_{2}\right)    =   C_{22}%
\overline{D}_{1}\overline{D}_{2}\left( F_{2}\right).
\end{gather*}
The above quantization map commutes with the action of $X_{x^{2}}$ if and
only if the coef\/f\/icients $C_{ij}$\ satisfy the following system of linear equations:
\begin{gather*}
2\left( \mu -\lambda -1\right) C_{11}   =   \left( -1\right) ^{p\left(
F\right) }, \\
2\left( \mu -\lambda -1\right) C_{12}   =   \left( -1\right) ^{p\left(
F\right) }\left( 1+2\lambda \right),  \\
2\left( \mu -\lambda -1\right) C_{13}   =   \left( -1\right) ^{p\left(
F\right) }, \\
2\left( \mu -\lambda -1\right) C_{14}   =   \left( -1\right) ^{p\left(
F\right)+1 }\left( 1+2\lambda \right),  \\
\left( \mu -\lambda -1\right) C_{21}   =   -\lambda,  \\
\left( 2\left( \mu -\lambda \right) -1\right) C_{22}   =   \left( -1\right)
^{p\left( F\right) }\lambda C_{12}-\lambda C_{14}.
\end{gather*}
Solving this system, one obtains the formula (\ref{FoQ1}).
\end{proof}

\subsection[Equivariant quantization map in the case of $3/2$-order
differential operators]{Equivariant quantization map in the case of $\boldsymbol{\frac{3}{2}}$-order
dif\/ferential operators}

Consider now the space of dif\/ferential operators $\mathcal{D}
_{\lambda \mu }^{\frac{3}{2}}\left( S^{1|2}\right) $.

\begin{proposition}
The unique $\mathrm{osp}\left( 2|2\right) $-equivariant quantization
map associates the following differential operator to a symbol
$\left( F_{1},F_{2}\right) \in \mathcal{S}_{\mu -\lambda
}^{\frac{3}{2}}\left( S^{1|2}\right) $:
\begin{gather}
Q\left( F_{1},F_{2}\right)   =   F_{1} \partial _{x}\overline{D}%
_{1}+F_{2} \partial _{x}\overline{D}_{2}
  +(-1)^{p(F)} \frac{\lambda +\frac{1}{2}}{\lambda -\mu +1}\left(
\overline{D}_{1}\left( F_{1}\right) +\overline{D}_{2}\left( F_{2}\right)
\right) \partial _{x} \nonumber\\
\phantom{Q\left( F_{1},F_{2}\right)   =}{}
+(-1)^{p(F)}\frac{1}{2(\lambda -\mu +1)}\left( \overline{D}%
_{2}\left( F_{1}\right) -\overline{D}_{1}\left( F_{2}\right) \right)
\overline{D}_{1}\overline{D}_{2} \nonumber\\
\phantom{Q\left( F_{1},F_{2}\right)   =}{}
 +\frac{(\lambda +\frac{1}{2})(\lambda -\mu +\frac{1}{2})}{(\lambda
-\mu +1)^{2}}\left( \partial _{x}\left( F_{1}\right) \overline{D}%
_{1}+\partial _{x}\left( F_{2}\right) \overline{D}_{2}\right) \nonumber\\
\phantom{Q\left( F_{1},F_{2}\right)   =}{}
 -\frac{\lambda +\frac{1}{2}}{2(\lambda -\mu +1)^{2}}\left( \overline{D%
}_{1}\overline{D}_{2}\left( F_{1}\right) \overline{D}_{2}-\overline{D}_{1}%
\overline{D}_{2}\left( F_{2}\right) \overline{D}_{1}\right) \nonumber\\
\phantom{Q\left( F_{1},F_{2}\right)   =}{}
 +(-1)^{p(F)} \frac{\lambda(\lambda +\frac{1}{2})}{(\lambda -\mu
+1)^{2}}\left( \partial _{x}\overline{D}_{1}\left( F_{1}\right) +\partial
_{x}\overline{D}_{2}\left( F_{2}\right) \right).
\label{FoQ32}
\end{gather}
\end{proposition}

\begin{proof}
An arbitrary $\mathrm{Af\/f}\left( 2|2\right) $-equivariant quantization map
is given by
\begin{gather*}
Q\left( F_{1},F_{2}\right)    =   F_{1}\partial _{x}\overline{D}%
_{1}+F_{2}\partial _{x}\overline{D}_{2}
+ \widetilde{Q}_{1}^{\left( 1\right) }\left( F_{1}\right) +\widetilde{Q}%
_{2}^{\left( 1\right) }\left( F_{2}\right)  \\
\phantom{Q\left( F_{1},F_{2}\right)    =}{}
+ \widetilde{Q}_{1}^{\left( 2\right) }\left( F_{1}\right) +\widetilde{Q}%
_{2}^{\left( 2\right) }\left( F_{2}\right)
+ \widetilde{Q}_{1}^{\left( 3\right) }\left( F_{1}\right) +\widetilde{Q}%
_{2}^{\left( 3\right) }\left( F_{2}\right),
\end{gather*}
where
\begin{gather*}
\widetilde{Q}_{1}^{\left( 1\right) }\left( F_{1}\right)    =   C_{11}%
\overline{D}_{1}\left( F_{1}\right) \partial _{x}+C_{13}\overline{D}%
_{2}\left( F_{1}\right) \overline{D}_{1}\overline{D}_{2}, \\
\widetilde{Q}_{1}^{\left( 2\right) }\left( F_{1}\right)    =
C_{21}\partial _{x}\left( F_{1}\right) \overline{D}_{1}+C_{23}\overline{D}%
_{1}\overline{D}_{2}\left( F_{1}\right) \overline{D}_{2}, \\
\widetilde{Q}_{1}^{\left( 3\right) }\left( F_{1}\right)    =
C_{31}\partial _{x}\overline{D}_{1}\left( F_{1}\right),  \\
\widetilde{Q}_{2}^{\left( 1\right) }\left( F_{2}\right)    =   C_{14}%
\overline{D}_{1}\left( F_{2}\right) \overline{D}_{1}\overline{D}_{2}+C_{12}%
\overline{D}_{2}\left( F_{2}\right) \partial _{x} ,\\
\widetilde{Q}_{2}^{\left( 2\right) }\left( F_{2}\right)    =
C_{24}\partial _{x}\left( F_{2}\right) \overline{D}_{2}+C_{22}\overline{D}%
_{1}\overline{D}_{2}\left( F_{2}\right) \overline{D}_{1}, \\
\widetilde{Q}_{2}^{\left( 3\right) }\left( F_{2}\right)    =
C_{32}\partial _{x}\overline{D}_{2}\left( F_{2}\right).
\end{gather*}
The above quantization map commutes with the action of $X_{x^{2}}$ if and
only if the coef\/f\/icients $C_{ij}$  satisfy the system of linear
equations:
\begin{gather*}
-2\left( \mu -\lambda -\tfrac{3}{2}\right) C_{11}-C_{12}  =  \left(
-1\right) ^{p\left( F\right) }\left( 1+2\lambda \right),  \\
-C_{11}-2\left( \mu -\lambda -\tfrac{3}{2}\right) C_{12}   =   \left(
-1\right) ^{p\left( F\right) }\left( 1+2\lambda \right),  \\
-2\left( \mu -\lambda -\tfrac{3}{2}\right) C_{13}+C_{14}   =   \left(
-1\right) ^{p\left( F\right) }, \\
C_{13}-2\left( \mu -\lambda -\tfrac{3}{2}\right) C_{14}   =   \left(
-1\right) ^{p\left( F\right)+1 }, \\
2\left( \mu -\lambda -\tfrac{3}{2}\right) C_{21}-C_{22}   =   -\left(
1+2\lambda \right),  \\
-C_{21}+4\left( \allowbreak \mu -\lambda -1\right) C_{22}   =
-C_{12}+\left( 1+2\lambda \right) C_{14}-C_{24}, \\
C_{23}+2\left( \mu -\lambda -\tfrac{3}{2}\right) C_{24}   =   -\left(
1+2\lambda \right),  \\
4\left( \allowbreak \mu -\lambda -1\right) C_{23}+C_{24}   =   C_{11}+\left(
1+2\lambda \right) C_{13}+C_{21}, \\
4\left( \lambda -\mu +1\right) C_{31}-C_{32}   =   \left( -1\right)
^{p\left( F\right) }2\lambda\left(C_{11}+C_{21}\right), \\
-C_{31}+4\left( \lambda -\mu +1\right) C_{32}  =  \left( -1\right)
^{p\left( F\right) }2\lambda \left(C_{12}+C_{24}\right).
\end{gather*}
Solving this system, one obtains the formula (\ref{FoQ32}).
\end{proof}

\subsection{Case of 2-contact order dif\/ferential operators}

The last case we consider is the space of dif\/ferential operators
$\mathcal{D}_{\lambda \mu }^{2}\left( S^{1|2}\right) $.
The proof of the following statement is again similar to that of
Proposition~\ref{FoDPr}; we will omit some details of calculations.

\begin{proposition}
The unique $\mathrm{osp}\left( 2|2\right) $-equivariant quantization map
associates the following differential operator to a symbol $\left(
F_{1},F_{2}\right) \in \mathcal{S}_{\mu -\lambda }^{2}\left( S^{1|2}\right) $:
\begin{gather}
Q\left( F_{1},F_{2}\right)  =  F_{1}\partial _{x}^{2}+F_{2}\partial _{x}
\overline{D}_{1}\overline{D}_{2}
   +(-1)^{p(F)} \frac{1}{\mu -\lambda -2}\left( \overline{D}_{1}\left(
F_{1}\right) \partial _{x}\overline{D}_{1}+\overline{D}_{2}\left(
F_{1}\right) \partial _{x}\overline{D}_{2}\right) \nonumber\\
 \phantom{Q\left( F_{1},F_{2}\right)  =}{}
   +(-1)^{p(F)} \frac{\lambda +1}{\mu -\lambda -2}\left( \overline{D}
_{2}\left( F_{2}\right) \partial _{x}\overline{D}_{1}-\overline{D}_{1}\left(
F_{2}\right) \partial _{x}\overline{D}_{2}\right) \nonumber\\
 \phantom{Q\left( F_{1},F_{2}\right)  =}{}  +\frac{2\lambda +1}{\lambda -\mu +2} \partial _{x}\left( F_{1}\right)
\partial _{x}+\frac{\lambda +1}{\lambda -\mu +2} \partial _{x}\left(
F_{2}\right) \overline{D}_{1}\overline{D}_{2} \nonumber\\
\phantom{Q\left( F_{1},F_{2}\right)  =}{}   -\frac{1}{\left( \lambda -\mu +2\right) \left( 2\lambda -2\mu +3\right)
} \overline{D}_{1}\overline{D}_{2}\left( F_{1}\right) \overline{D}_{1}
\overline{D}_{2} \nonumber\\
 \phantom{Q\left( F_{1},F_{2}\right)  =}{}
   +\frac{(2\lambda +1) (\lambda +1)}{\left( \lambda -\mu +2\right)
\left( 2\lambda -2\mu +3\right) } \overline{D}_{1}\overline{D}_{2}\left(
F_{2}\right) \partial _{x} \nonumber\\
\phantom{Q\left( F_{1},F_{2}\right)  =}{}
   -(-1)^{p(F)} \frac{2\lambda +1}{\left( \lambda -\mu +2\right) \left(
2\lambda -2\mu +3\right) }\left( \partial _{x}\overline{D}_{1}\left(
F_{1}\right) \overline{D}_{1}+\partial _{x}\overline{D}_{2}\left(
F_{1}\right) \overline{D}_{2}\right) \nonumber\\
\phantom{Q\left( F_{1},F_{2}\right)  =}{}
   -(-1)^{p(F)} \frac{(2\lambda +1) (\lambda +1)}{\left( \lambda -\mu
+2\right) \left( 2\lambda -2\mu +3\right) } \left( \partial _{x}\overline{D}
_{2}\left( F_{2}\right) \overline{D}_{1}-\partial _{x}\overline{D}_{1}\left(
F_{2}\right) \overline{D}_{2}\right) \nonumber\\
\phantom{Q\left( F_{1},F_{2}\right)  =}{}
   + \frac{\lambda  (2\lambda +1)}{\left( \lambda -\mu +2\right) \left(
2\lambda -2\mu +3\right) } \partial _{x}^{2}\left( F_{1}\right) \nonumber\\
 \phantom{Q\left( F_{1},F_{2}\right)  =}{}
   +\frac{\lambda  (2\lambda +1)(\lambda +1)}{\left( \lambda -\mu
+2\right) \left( 2\lambda -2\mu +3\right) \left( \lambda -\mu +1\right) }
\partial _{x}\overline{D}_{1}\overline{D}_{2}\left( F_{2}\right),\label{LastEq}
\end{gather}
provided $\mu -\lambda \neq 1,\frac{3}{2},2$.
\end{proposition}

\begin{proof}
An arbitrary $\mathrm{Af\/f}\left( 2|2\right) $-equivariant
quantization map is given by
\begin{gather*}
Q\left( F_{1},F_{2}\right)   =  F_{1}\partial _{x}^{2}+F_{2}\partial _{x}%
\overline{D}_{1}\overline{D}_{2}
+ \widetilde{Q}_{1}^{\left( 1\right) }\left( F_{1}\right) +\widetilde{Q}%
_{2}^{\left( 1\right) }\left( F_{2}\right)  \\
\phantom{Q\left( F_{1},F_{2}\right)   =}{}+
 \widetilde{Q}_{1}^{\left( 2\right) }\left( F_{1}\right) +\widetilde{Q}%
_{2}^{\left( 2\right) }\left( F_{2}\right)
+ \widetilde{Q}_{1}^{\left( 3\right) }\left( F_{1}\right) +\widetilde{Q}%
_{2}^{\left( 3\right) }\left( F_{2}\right)
+ \widetilde{Q}_{1}^{\left( 4\right) }\left( F_{1}\right) +\widetilde{Q}%
_{2}^{\left( 4\right) }\left( F_{2}\right),
\end{gather*}
where
\begin{gather*}
\widetilde{Q}_{1}^{\left( 1\right) }\left( F_{1}\right)    =   C_{11}%
\overline{D}_{1}\left( F_{1}\right) \partial _{x}\overline{D}_{1}+C_{13}%
\overline{D}_{2}\left( F_{1}\right) \partial _{x}\overline{D}_{2}, \\
\widetilde{Q}_{1}^{\left( 2\right) }\left( F_{1}\right)    =
C_{21}\partial _{x}\left( F_{1}\right) \partial _{x}+C_{23}\overline{D}_{1}%
\overline{D}_{2}\left( F_{1}\right) \overline{D}_{1}\overline{D}_{2}, \\
\widetilde{Q}_{1}^{\left( 3\right) }\left( F_{1}\right)    =
C_{31}\partial _{x}\overline{D}_{1}\left( F_{1}\right) \overline{D}%
_{1}+C_{33}\partial _{x}\overline{D}_{2}\left( F_{1}\right) \overline{D}_{2},
\\
\widetilde{Q}_{1}^{\left( 4\right) }\left( F_{1}\right)    =
C_{41}\partial _{x}^{2}\left( F_{1}\right),  \\
\widetilde{Q}_{2}^{\left( 1\right) }\left( F_{2}\right)    =   C_{14}%
\overline{D}_{1}\left( F_{2}\right) \partial _{x}\overline{D}_{2}+C_{12}%
\overline{D}_{2}\left( F_{2}\right) \partial _{x}\overline{D}_{1}, \\
\widetilde{Q}_{2}^{\left( 2\right) }\left( F_{2}\right)    =
C_{24}\partial _{x}\left( F_{2}\right) \overline{D}_{1}\overline{D}%
_{2}+C_{22}\overline{D}_{1}\overline{D}_{2}\left( F_{2}\right) \partial _{x},
\\
\widetilde{Q}_{2}^{\left( 3\right) }\left( F_{2}\right)    =
C_{34}\partial _{x}\overline{D}_{1}\left( F_{2}\right) \overline{D}%
_{2}+C_{32}\partial _{x}\overline{D}_{2}\left( F_{2}\right) \overline{D}_{1},
\\
\widetilde{Q}_{2}^{\left( 4\right) }\left( F_{2}\right)    =
C_{42}\partial _{x}\overline{D}_{1}\overline{D}_{2}\left( F_{2}\right).
\end{gather*}
The above quantization map commutes with the action of $X_{x^2}$
and therefore is $\mathrm{osp}\left( 2|2\right) $-equivariant
if and only if the coef\/f\/icients $C_{ij}$ are as in (\ref{LastEq}).
\end{proof}

\subsection*{Acknowledgements} I am grateful to H.~Gargoubi and V.~Ovsienko for
the statement of the problem and constant help. I am also pleased to thank
D.~Leites for critical reading of this paper and a number of helpful
suggestions.

\pdfbookmark[1]{References}{ref}
\LastPageEnding

\end{document}